\documentclass[aps,prd,preprint,tightenlines,groupedaddress,showpacs]{revtex4}
\usepackage{epsf,epsfig,graphics,graphicx}
\usepackage{graphicx}

\usepackage{dcolumn}
\usepackage{bm}
\usepackage{amsmath}
\newcommand{\mb}{\bar{\rho}_{\rm{m}}}
\newcommand{\mc}{\rho_{\rm{m}}}
\newcommand{\deb}{\bar{\rho}_{\rm{de}}}
\newcommand{\dec}{\rho_{\rm{de}}}
\newcommand{\m}{{\rm m}}
\newcommand{\de}{{\rm de}}
\newcommand{\D}{\Delta}
\newcommand{\ta}{{\rm ta}}
\newcommand{\vir}{{\rm vir}}
\newcommand{\Om}{\Omega}
\newcommand{\f}{\frac}
\newcommand{\nld}{\delta^{\rm{NL}}}
\newcommand{\dm}{\delta_{\m}}
\newcommand{\dde}{\delta_{\de}}
\newcommand{\x}{\tilde{x}}
\newcommand{\y}{\tilde{y}}
\newcommand{\dmi}{\delta_{\rm{m},\it{i}}}
\newcommand{\ddei}{\delta_{\rm{de},\it{i}}}
\newcommand{\dc}{\delta_{\rm{c}}}
\newcommand{\dvir}{\delta_\vir}
\newcommand{\Dvir}{\Delta_\vir}
\newcommand{\EdS}{{\rm EdS}}
\newcommand{\etac}{\eta_{\rm c}}

\begin{document}
\title{Spherical Collapse Models with Clustered Dark Energy}
\author{Chia-Chun Chang$^1$}
\author{Wolung Lee$^1$}
\author{Kin-Wang Ng$^{2,3}$}
\email{leewl@phy.ntnu.edu.tw}
\affiliation{
$^1$Department of Physics, National Taiwan Normal University,
Taipei 11677, Taiwan\\
$^2$Institute of Physics, Academia Sinica, Taipei 11529, Taiwan\\
$^3$Institute of Astronomy and Astrophysics, Academia Sinica, Taipei 11529, Taiwan}

\date{\today}
\begin{abstract}

We investigate the clustering effect of dark energy (DE) in the formation of galaxy clusters using the spherical collapse model. Assuming a fully clustered DE component, the spherical overdense region is treated as an isolated system which conserves the energy separately for both matter and DE inside the spherical region. Then, by introducing a parameter $r$ to characterize the degree of DE clustering, which is defined by the nonlinear density contrast ratio of matter to DE at turnaround in the recollapsing process, i.e. $r\equiv \nld_{\de,\ta}/\nld_{\m,\ta}$, we are able to uniquely determine the spherical collapsing process and hence obtain the virialized overdensity $\Dvir$ through a proper virialization scheme. Estimation of the virialized overdensities from current observation on galaxy clusters suggests that $0.5 < r < 0.8$ at $1\sigma$ level for the clustered DE with $w < -0.9$.  Also, we compare our method to the linear perturbation theory that deals with the growth of DE perturbation at early times. While both results are consistent with each other, our method is practically simple and it shows that the collapse process is rather independent of initial DE perturbation and its evolution at early times.

\end{abstract}

\pacs{98.80.-k, 98.65.-r}
\maketitle
\section{Introduction}
The modern integrated analysis of observational evidences from Type Ia supernovae at high redshifts, the cosmic microwave background (CMB), and various surveys on the large scale structure (LSS) has concordantly indicated the accelerating expansion of our Universe~\cite{obs}. Though its very nature remains elusive, the dark energy (DE) which contributes a strong negative pressure at large scales provides a tantalizing explanation for such an unexpected gravitating behavior.

From the perspective of structure formations, the spherical collapse model (SCM) first introduced by Gunn and Gott in 1972~\cite{gg} is a simple yet effective method in analyzing the nonlinear evolution of a spherical overdensity in the Universe via three consecutive phases during which the matter overdensity would (1) expand to the maximum size along with the background universe, then (2) turn around under its own gravity, and finally (3) collapse to form large scale structures. Apparently, the SCM is a practical tool readily for exploring how DE plays a role in the formation of cosmic structures.

In the context of the SCM, the DE component that drives the cosmic acceleration is generically classified into two categories. In the first category, in most cases the DE in the overdense spherical region is simply treated as a part of the homogeneous DE background without coupling to any matter components. As such, the overdense sphere is indeed an open system in which the total energy is not conserved~\cite{sky,pace,horellou,bartel,meyer,percival,pwang06,naderi}. On the other hand, certain models of structure formation~\cite{crem,yywong,voglis,edebp,nunesmota,abramo07} allow some forms of DE-matter coupling. As a consequence, the overdense region is segregated from the expanding background so that the total energy conserves within such a spherically isolated system throughout the entire collapsing course. These two distinctive scenarios of structure formation, leading to different properties of the large-scale structures, can provide an observational test of the DE models.

Here we scrutinize the scenario in which the DE completely clusters with matter components and the energy conserves within a spherically isolated overdensity. However, within the context of the SCM, the system is under-determined due to the fact that the relative magnitude of the DE density to the matter density within the overdense region is unknown. Since the exact property is unknown, particularly at early times in the cosmic history, the DE is usually modeled by various types of scalar fields~\cite{crem,nunesmota,motabruck}. Given a scalar field model for the DE, one can trace the evolution of DE perturbation to find the ratio of DE and matter densities at any time. But one needs to be cautious about a caveat that the scalar field model, giving rise to a specific equation of state, may oversimplify the property of the DE. For instance, the DE may behave quite differently in late times in the early DE models~\cite{earlyDE}, or if there involves a phase transition in the scalar field that changes its equation of state. As long as being well gauged by observational constraints on the formation of structures in relatively late times, it is sufficient enough to mimic the evolution of DE by its equation of state without resorting to a specific scalar field model. Accordingly, we will retain macroscopic physical quantities, namely, pressure and energy densities of all relevant components as basic degrees of freedom to decipher the possible clustering effect from the DE component within the SCM. The drawback is that we will need to introduce a new degree of freedom to parameterize the unknown DE density inside the collapsing overdense region. We will show that this can be done in a very simple and self-consistent way.

One further advantage of the SCM is that it enables us to differentiate a fully clustered DE from the cosmological constant in virtue of the virial theorem. As being well known that a proper virialization is necessary to form a stable structure without involving an unreasonable disaster of singularity~\cite{sky,lahav}. Hence we presuppose that the whole collapsing system, including the DE component that drives the present cosmic acceleration, is virialized in the SCM analysis. However, if the cosmological constant is responsible for the current acceleration of the Universe, it would not participate in the virialization due to the nature of a constant. Subsequently, one is able to discriminate the model with a fully clustered DE from that with a cosmological constant.

This paper is organized as follows: For the nonlinear evolution of matter overdensity, we numerically solve the evolution equations of the overdense region in Sec.~\uppercase\expandafter{\romannumeral 2} and compute the virialized overdensity $\D_{\vir}$ in Sec.~\uppercase\expandafter{\romannumeral 3}. In Sec.~\uppercase\expandafter{\romannumeral 4}, combining the SCM with the linear evolution of DE and matter density perturbation at early times, we track down the complete evolution of the clustered DE and matter. Then, we show that this is equivalent to introducing the new parameter as mentioned above. Furthermore, we can use observational data of galaxy clusters to constrain the parameter. Finally, we summarize our conclusions in Sec.~\uppercase\expandafter{\romannumeral 5}.

\section{Spherical Collapsing with clustered Dark Energy}

The evolution of a spatially flat universe with a homogeneous and isotropic background is usually governed by
\begin{eqnarray}
\label{H}
& & H^2=\left(\frac{\dot{a}}{a}\right)^2=\frac{8\pi G}{3}(\mb+\deb),\\
\label{a}
& & \frac{\ddot{a}}{a}=-\f{4\pi G}{3}\left[\mb+(1+3w)\deb\right],\\
\label{mb}
& & \dot{\bar{\rho}}_{\m}+3\left(\frac{\dot{a}}{a}\right)\mb=0,\\
\label{deb}
& & \dot{\bar{\rho}}_{\de}+3(1+w)\left(\frac{\dot{a}}{a}\right)\deb=0,
\end{eqnarray}
where the overdot denotes the derivative with respect to time $t$; $ a $ is the scale factor in the Friedmann-Lema\^{i}tre-Robertson-Walker (FLRW) metric; $ \mb $ and $ \deb $ denote respectively the energy densities of matter and DE.  We have ignored the radiation component because it plays an irrelevant role in structure collapse. Also, we have assumed a constant equation of state (eos) of the DE, i.e. $ w\equiv\bar{P}_{\de}/\deb $.

As a first step to understand the large-scale structures in such a universe, we explore the nonlinear gravitational collapse in matter using the simple spherical collapse model~\cite{gg}. We consider a spherically symmetric overdense region with both $ \mc $ and $ \dec $ inside sitting on top of the otherwise uniform background. Given a sufficiently large initial density contrast in the matter sector, the overdense region expands with the background as the cosmic time unfolds. It then turns around under the influence of its own gravity after reaching the critical size and collapses to form structures. The evolution of the overdensity of radius $ R $ can be described by the dynamical equations similar to those governing the motion of the background universe,

\begin{eqnarray}
\label{R}
& & \frac{\ddot{R}}{R}=-\frac{4\pi G}{3}\left[\mc+(1+3w)\dec\right],\\
\label{mc}
& & \dot{\rho}_{\m}+3\left(\frac{\dot{R}}{R}\right)\mc=0,\\
\label{dec}
& & \dot{\rho}_{\de}+3(1+w)\left(\frac{\dot{R}}{R}\right)\dec
=\alpha\Gamma,  
\end{eqnarray}
where
\[\Gamma=3(1+w)\left(\frac{\dot{R}}{R}-\frac{\dot{a}}{a}\right)\dec,~~~
\mbox{with}~~~0\le\alpha\le1.
\]
The DE component within the spherical overdense region assumes the same eos as that in the background universe, i.e., $ P_{\de}=w\dec $. The parameter $\alpha$ controls the energy balance in the DE sector which is characterized by $\Gamma$. For the non-clustering case, $ \alpha=1 $, one is incapable of differentiating the behavior of DE component inside and outside of the spherical overdensity. Under the circumstance, $ \dec=\deb $, and the energy does not conserve within the overdensity~\cite{sky,motabruck}. On the contrary, the fully clustering case with $ \alpha=0 $ renders $ \dec \neq \deb$ such that the spherical overdense region is effectively segregated from the background universe and is considered an isolated system satisfying the law of energy conservation.

To solve the model numerically, we recast $ a $ and $ R $ in terms of their corresponding values at the moment of turnaround such that 

\begin{equation}
x\equiv\frac{a}{a_{\ta}},
\end{equation}

\begin{equation}
y\equiv\frac{R}{R_{\ta}}.
\end{equation}
With the help of Eqs. \eqref{mc} and \eqref{dec}, Eqs. \eqref{H} and  \eqref{R} become

\begin{eqnarray}
\label{dx} 
\frac{dx}{d\tau}      &=& \sqrt{x^{-1}+\frac{1}{Q_{\ta}}x^{-3w-1}},\\
\label{dy}
\frac{d^2 y}{d\tau^2} &=& -\frac{1}{2}\left[
\zeta y^{-2}+(1+3w)\frac{1}{Q_{\rm{cta}}}x^{-3(1+w)\alpha}
y^{-3(1+w)(1-\alpha)+1}
\right],
\end{eqnarray}
where
\begin{equation*}
d\tau\equiv H_{\ta}\sqrt{\Om_{\m}(x_{\ta})}dt , ~~
\zeta\equiv\left.\f{\mc}{\mb}\right|_{z_{\ta}}\hspace{-1mm} ,~~
Q_{\ta}\equiv\left.\f{\mb}{\deb}\right|_{z_{\ta}}\hspace{-2mm}=\f{\Om_{\m,0}}{\Om_{\de,0}}(1+z_{\ta})^{-3w}\hspace{-1mm}, ~~{\rm and}~~
Q_{\rm{cta}}\equiv\left.\f{\mb}{\dec}\right|_{z_{\ta}}\hspace{-2mm}.
\end{equation*}
%
%
\begin{figure}[htb]
\begin{minipage}[b]{0.7\linewidth}
\centering
\includegraphics[width=\textwidth]{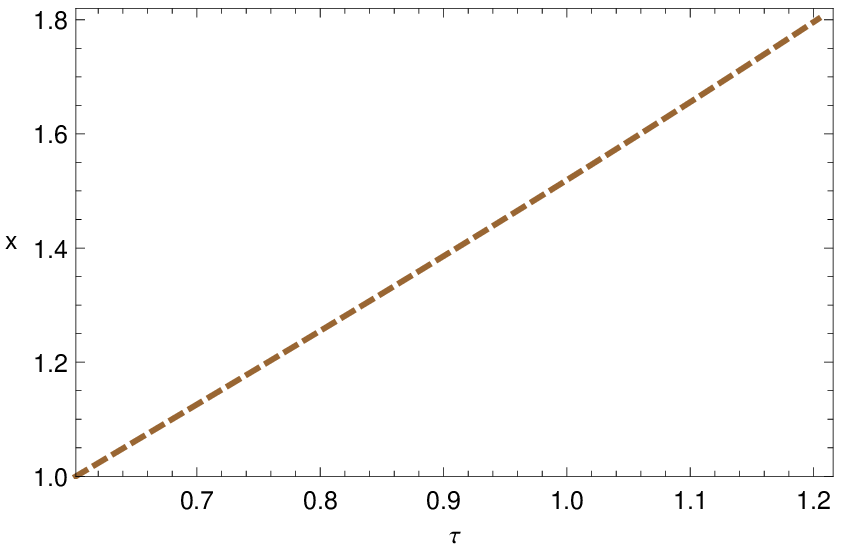}
\end{minipage}
\begin{minipage}[b]{0.7\linewidth}
\centering
\includegraphics[width=\textwidth]{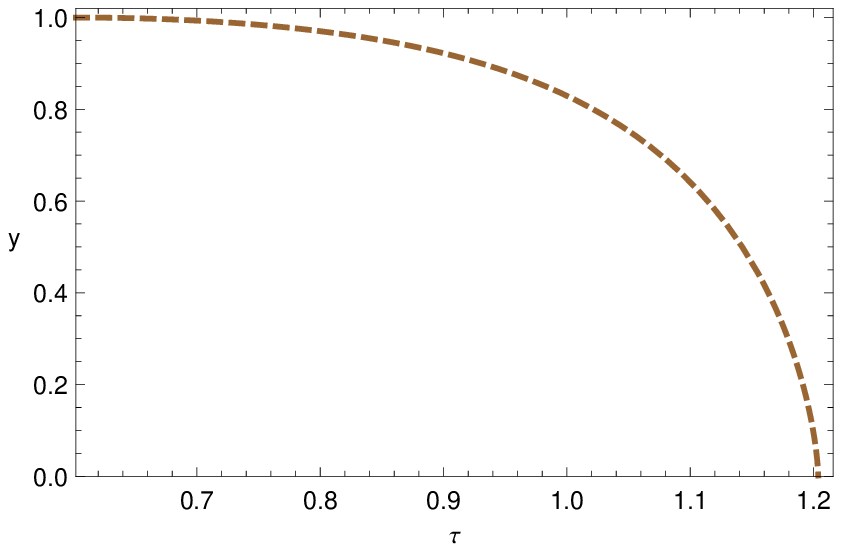}
\end{minipage}
\caption{The numerical solutions to Eqs.~\eqref{dx} and \eqref{dy_cls} for $ w=-0.8 $,
$ z_{\ta}=0.8 $, and $ r=1 $. The upper panel shows the evolution of the background universe since the turnaround time. The lower panel demonstrates the post-turnaround evolution of the spherical overdensity.}
\label{fig:fig1}
\end{figure}
%
%
\begin{figure}[htb]
\includegraphics[width=0.7\textwidth]{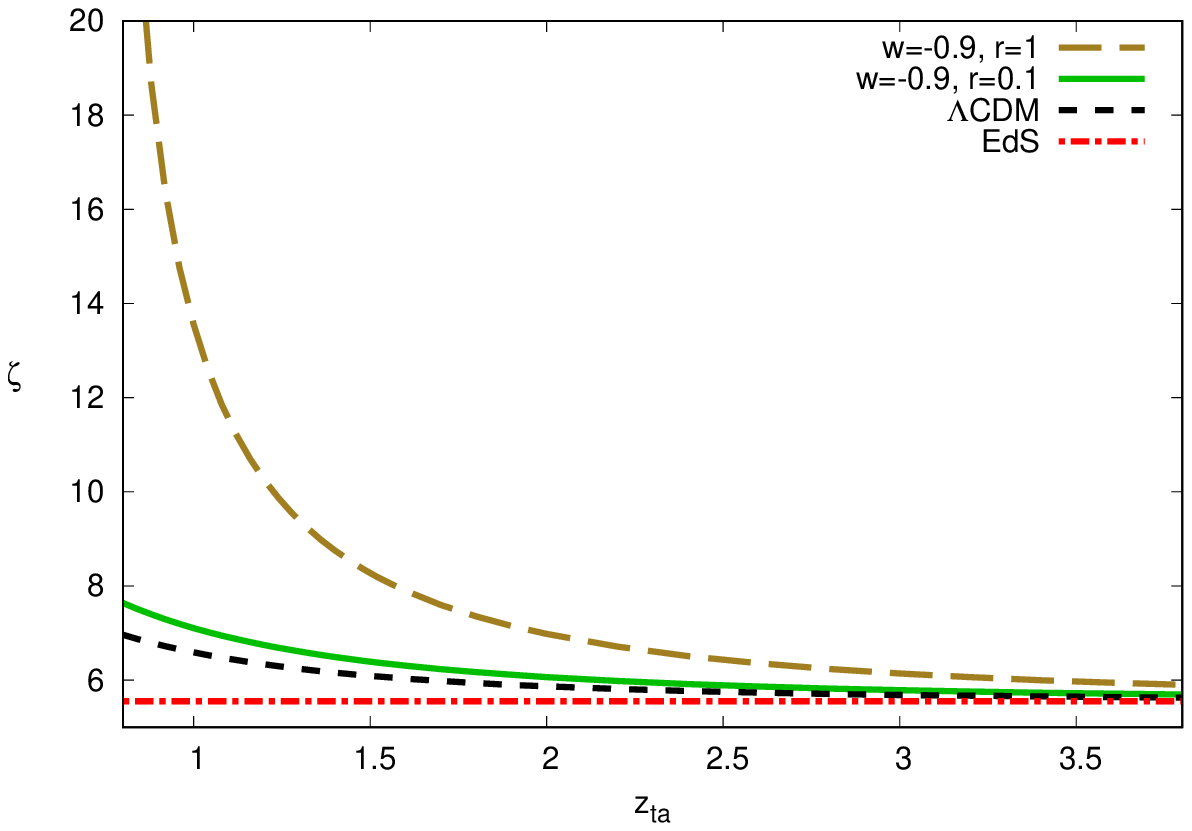}
\caption{The overdensity $ \zeta $ at turnaround as a function of $ z_{\ta} $ for a couple of DE models with the same eos $w$ but different $r$. The cases for the standard $ \Lambda\rm{CDM} $, which carries no dependence on $ r $, and the Einstein-de Sitter universe where $\zeta_{\rm EdS}=5.55$ are also plotted.}
\label{fig:fig2}
\end{figure}
The turnaround time $\tau_{\ta}$, signified as $x_{\ta}=1$, can be determined by means of the hypergeometric function $F$ in terms of the eos $w$ and the turnaround redshift $z_{\ta}$ as~\cite{sky}

\begin{equation}
\label{tta}
\tau_{\ta}
=\f{2}{3}F\left[\f{1}{2}, -\f{1}{2w}, 1-\f{1}{2w}, -Q_{\ta}^{-1}\right].
\end{equation}
The present values of density parameters in the matter and DE components are respectively specified by $ \Om_{\m,0} $ and $ \Om_{\de,0}$. Consequently, Eqs. \eqref{dx} and \eqref{dy} are suitable for mimicking the evolution of overdensity during the recollapsing process.

For the non-clustering case ($\alpha=1$) where the behavior of DE is indistinguishable inside and outside of the overdense region, $Q_{\rm{cta}}=Q_{\ta}$, and the solutions are well discussed in Ref.~\cite{sky}. The Einstein-de Sitter (EdS) universe involving no DE is considered as a special example of the non-clustering case. Since $\Om_{\m}^0=1$ and $(Q_{\rm{cta}}^{\rm EdS})^{-1}=(Q_{\ta}^{\rm EdS})^{-1}=0$ for the EdS model, Eq.~\eqref{dy} reduces to

\begin{equation}
\f{dy^2}{d\tau^2}=-\f{\zeta_\EdS}{2y^2},
\nonumber
\end{equation}
where the overdensity at turnaround $\zeta_\EdS=9\pi^2/16\approx5.55$ is a well-known value for this matter-dominated universe.

For the clustering case with $\alpha=0$, Eq. \eqref{dy} can be further simplified as

\begin{equation}
\label{dy_cls}
\f{d^2y}{d\tau^2}=-\f{1}{2}\left[
\zeta y^{-2}+\left(\f{1+3w}{Q_{\rm{cta}}}\right)y^{-2-3w}
\right].
\end{equation}
The scenario of SCM requires that the overdense region shall directly collapse to $R=0$ at the end. Because of the time reversal symmetry of Eq.~\eqref{dy_cls}, the evolution of the overdense region should be symmetric before and after the turnaround epoch. As a result, given $w$ and $z_{\ta}$, it is straightforward to locate a proper value for $ \zeta $ by fulfilling the criterion that $y(\tau_{\rm f})=y(2\tau_{\ta})=0$ while solving Eq. \eqref{dy_cls} numerically, in which $\tau_{\rm f}$ represents the moment of complete collapse.

A justification is in order, however. According to Eq. \eqref{dy_cls}, cosmic structures are supposed to form at the end of the spherical collapsing process , but it does not necessarily imply that any matter density contrast would turn around and collapse at a certain epoch. As a matter of fact, there are many run-away solutions without collapsing to Eq. \eqref{dy_cls} as the magnitudes of the initial matter density contrast $ \dmi $ are inadequate to form structures. We will scrutinize the linear evolution of $\dm$ in the context of SCM later in Sec.~\uppercase\expandafter{\romannumeral 4}. For the time being, let us focus only on the nonlinear collapsing process of matter density perturbation. We thus look for solutions to Eq.~\eqref{dy_cls} that fulfill the boundary conditions, $ y'(\tau_{\ta})=0 $ and $ y(\tau_{\rm f})=0 $. These solutions do exist and each solution corresponds an initial $\dm$ that is guaranteed to turn around at the moment $\tau_{\ta}$ then collapse completely to form structures at $\tau_{\rm f}=2\tau_{\ta}$.

For models with $\alpha=0$, however, the DE density within the overdense region is unknown because that $ \dec\neq\deb $, i.e. $ Q_{\rm{cta}} $ is undetermined in Eq. \eqref{dy_cls}. To resolve the difficulty we assume a simple linear relation between the nonlinear density contrasts in the DE and the matter sectors at turnaround with a parameter $r$ satisfying

\begin{equation}
\label{r}
\nld_{\de,\ta}=r\nld_{\m,\ta}~,
\end{equation}
where  $ \nld_{\m}\equiv\mc/\mb-1 $, and  $ \nld_{\de}\equiv\dec/\deb-1 $.
Accordingly, $ Q_{\rm{cta}} $ can be expressed in terms of $Q_{\ta}$ as
%
\begin{equation}
Q_{\rm{cta}}=\f{Q_{\ta}}{1+r[\zeta(w,z_{\ta},r)-1]}.
\end{equation}
Given appropriate values to $ w $, $ z_{\ta} $ and $ r $, Eq. \eqref{dy_cls} can now be solved completely.
As an example, Fig. 1 depicts the evolutionary trajectory of a fully clustered DE model with $ w=-0.8 $, $ z_{\ta}=0.8 $,
and $ r=1 $. We show the matter overdensity $\zeta$ at turnaround versus $z_{\ta}$ for a couple of clustered DE models, the $ \Lambda\rm{CDM} $, and the EdS universe in Fig. 2. Apparently, $\zeta$ is a monotonically decreasing function of turnaround redshift $z_{\ta}$ for DE models, including $\Lambda$CDM: the later a spherical overdense region turnarounds, the more matter density at the turning point it contains. The growth in $\zeta$ at the same $z_{\ta}$ is proportional to the eos $w$, which is expected owing to the weaker repelling effect exercised by DE models with a greater $w$. Moreover, the surge in amplitude of $\zeta$ increases drastically at later times as the ratio of the nonlinear DE density contrast at turnaround, i.e. the parameter $r$, increases. On the other hand, $\zeta$ approaches the limit of $\zeta_{\rm EdS}=5.55$ in all models of DE (including the case of $ \Lambda\rm{CDM} $) as the turnaround redshift becomes higher and higher.

\section{Virialization}

In reality, a spherically overdense region can only collapse to a certain finite extension $ R_{\rm{vir}} $ and virializes over a mass
$ M_{\rm{vir}} $ rather than reaching an unphysical singularity at $R=0$. By means of the virial theorem that $ T_{\rm{vir}}=(R \partial U/\partial R)_{\rm{vir}}/2$, the virialized radius $ R_{\rm{vir}} $ is established to be compatible with the kinetic energy $ T $ and the potential energy $ U $ of the fluid system~\cite{lahav}. For the clustering cases where $ \alpha=0 $ the energy conservation in the spherical overdensity renders
%
\begin{equation}
\label{virial}
\left[U+\f{R}{2}\f{\partial U}{\partial R}\right]_{\rm{vir}}
=U_{\ta}\,.
\end{equation}
Assuming that the whole system, including the matter and the DE, virializes within the boundary of the overdense region, the potential energy $ U $ of the system is obtained as
%
\begin{equation}
\label{U}
U=\f{1}{2}\int_0^R\mc\Phi_{\m}dV+\f{1}{2}\int_0^R\dec\Phi_{\m}dV
   +\f{1}{2}\int_0^R\mc\Phi_{\de}dV+\f{1}{2}\int_0^R\dec\Phi_{\de}dV.
\end{equation}
The potentials induced respectively by the matter and the DE components, i.e. $\Phi_{\m}$ and $\Phi_{\de}$, satisfy the general relation that
%
\begin{equation*}
\Phi_{\rm{x}}(\lambda)=-2\pi G (1+3w_{\rm{x}})\rho_{\rm{x}}
                     \left(R^2-\f{\lambda^2}{3}\right)
\end{equation*}
with the corresponding eos $ w_{\rm{x}} $ inside a uniform sphere of radius $ R $, which amounts to the size of the overdense region. Therefore, the matter-induced potential is practically attractive, i.e. $\Phi_{\m}<0$ since $w_{\m}=0$, whereas the DE-induced potential is efficently repelling, i.e. $\Phi_{\de}>0$ due to $w<-1/3$.

To measure the size of a virialized structure, one combines Eqs.~\eqref{virial} and~\eqref{U} to obtain
%
\begin{equation}
\label{eq_yvir}
 \left[1+(2+3w)q+(1+3w)q^2\right]y_{\vir} 
 -{(2+3w)(1-3w)\over 2}qy_{\vir}^{-3w}
    -{(1+3w)(1-6w)\over 2}q^2y_{\vir}^{-6w}=\f{1}{2}\, ,
\end{equation}
where $ q\equiv(\dec/\mc)|_{z_{\ta}} $ denotes the density ratio of DE to matter within the overdense region at turnaround. Using the background solutions,
$\bar{\rho}_{\m,\ta} = \bar{\rho}_{\m,0}(1+z_{\ta})^3$ and $\bar{\rho}_{\de,\ta} = \bar{\rho}_{\de,0}(1+z_{\ta})^{3(1+w)}$,
and the relation between their nonlinear density contrasts described in Eq.~\eqref{r}, the parameter $q$ can be rewritten as
\begin{equation}
\label{q}
q = \f{r(\zeta-1)+1}{\zeta} \left(\f{1-\Omega_{\m,0}}{\Omega_{\m,0}}\right)(1+z_{\ta})^{3w},
\end{equation}
where the overdensity at turnaround is characterized by
\begin{equation}
\zeta=1+\nld_{\m,\ta}~,
\end{equation}
according to its definition. Once $ w $, $ z_{\ta} $, and $ r $ are all specified, $ \zeta $ can be found, and so does $ q $. Subsequently, the size $ y_{\vir}$ of a completely virialized recollapsing system can be determined by the constraint Eq. \eqref{eq_yvir}.

For a system that is merely virialized in part, however, the equation of motion for $y_{\vir}$ should be properly modified. In the particular case of $\Lambda$CDM where only the matter component virializes within the boundary of the overdense region, the dark energy potential $\Phi_{\rm de}$ in Eq.~\eqref{U} is ineffective. Thus, the nonlinear perturbation in the DE is negligible (i.e., $r=0$), and Eq. \eqref{eq_yvir} is reduced to~\cite{lahav}
%
\begin{equation}
\label{eq_yvir_m}
(1+q)y_{\vir}-\f{q}{2}(1-3w)y_{\vir}^{-3w}=\f{1}{2}\,.
\end{equation}
On the other hand, the virialization in the mater-dominated EdS model is exclusively attributed to the first term in Eq.~\eqref{U}. It turns out that $y_{\vir}=1/2$, completely irrelevant to the turnaround redshift $ z_{\ta} $ in such a single component  universe.

\begin{figure}[htb]
\includegraphics[width=0.7\textwidth]{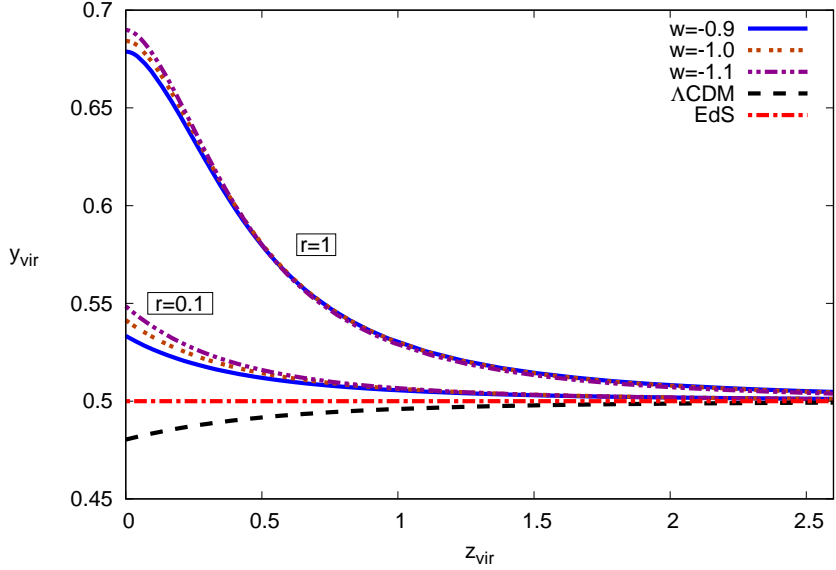}
\caption{The size of the recollapsing overdense region $y_{\vir}$ is plotted against the redshift at the epoch of virialization $z_{\vir}$ for the $\Lambda$CDM, and DE models with $w=-0.9,-1,-1.1$ for a fixed value of $r$ (indicated in a box).  The constant line of $y_{\vir}=0.5$ in the EdS model is also plotted for reference. The $\Lambda$CDM rendering $ y_{\vir}<0.5 $ is clearly distinguishable from other DE models with $w=-1$ but $r\neq 0$.}
\end{figure}

The relations between $ y_{\vir} $ and $z_{\vir}$ are shown in Fig. 3 for a couple of DE models with different values for $r$, and the $\Lambda$CDM, where the redshift at the epoch of virialization is determined as
\begin{equation}
z_{\vir}=\f{1+z_{\ta}}{x_{\vir}}-1,
\end{equation}
according to the definitions of redshift $z$ and the rescaled comoving scale factor $x$.  The value of $ x_{\vir} $ is easily derivable  from Eq.~\eqref{dx} once the final moment of the virialization process $\tau_{\vir}$ is determined from the solution of Eq. \eqref{dy_cls} for a specific value of $ y_{\vir} $.

Apparently, differences in the virialization condition enable us to distinguish the DE model with $w=-1$ from the $\Lambda$CDM. In addition, DE models can be properly classified into groups with respect to different values of $r$. Because the degree of inhomogeneity in DE increases with $r$, thus providing more negative pressure, a larger value in $r$ gives rise to a larger $y_{\vir}$, as illustrated in Fig. 3. On the other hand, within each $r$-group, the DE model with a larger $ w $ exerts less negative pressure upon the boundary of the overdense region and thus leading to a smaller $ y_{\vir} $. Moreover, all DE models approach the EdS universe in the limit of large $z_{\vir}$, regardless of how $r$ and $w$ may have varied. This trend simply reflects the feature of the matter domination at earlier times at which a DE model is basically indistinguishable from the EdS universe. Here we have assumed a constant $w$. This conclusion is not necessarily true when $w$ changes its value at early time such that a non-negligible amount of dark energy is present in the early time, for example, in the early dark energy models~\cite{earlyDE}.
\begin{figure}[htb]
\begin{minipage}[h]{0.7\linewidth}
\centering
\includegraphics[width=\textwidth]{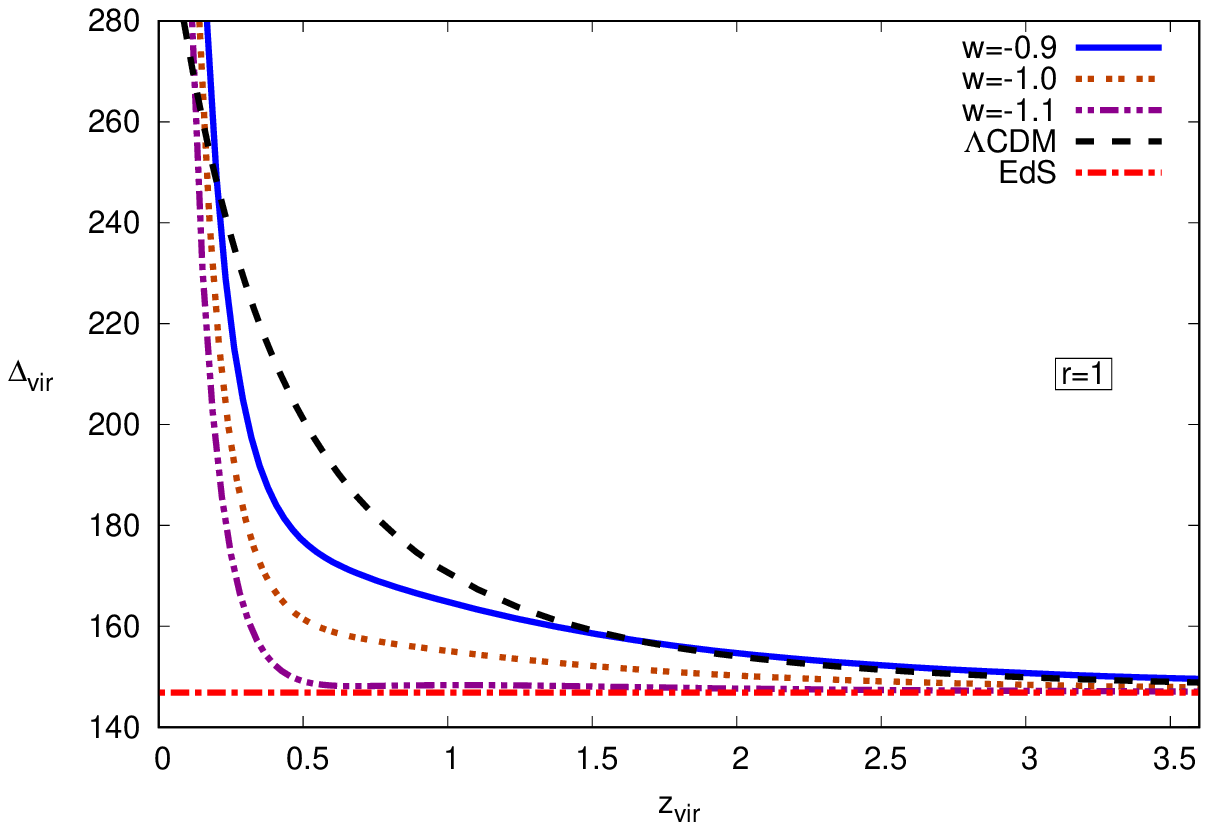}
\label{fig:figure1}
\end{minipage}
\hspace{0.5cm}
\begin{minipage}[h]{0.7\linewidth}
\centering
\includegraphics[width=\textwidth]{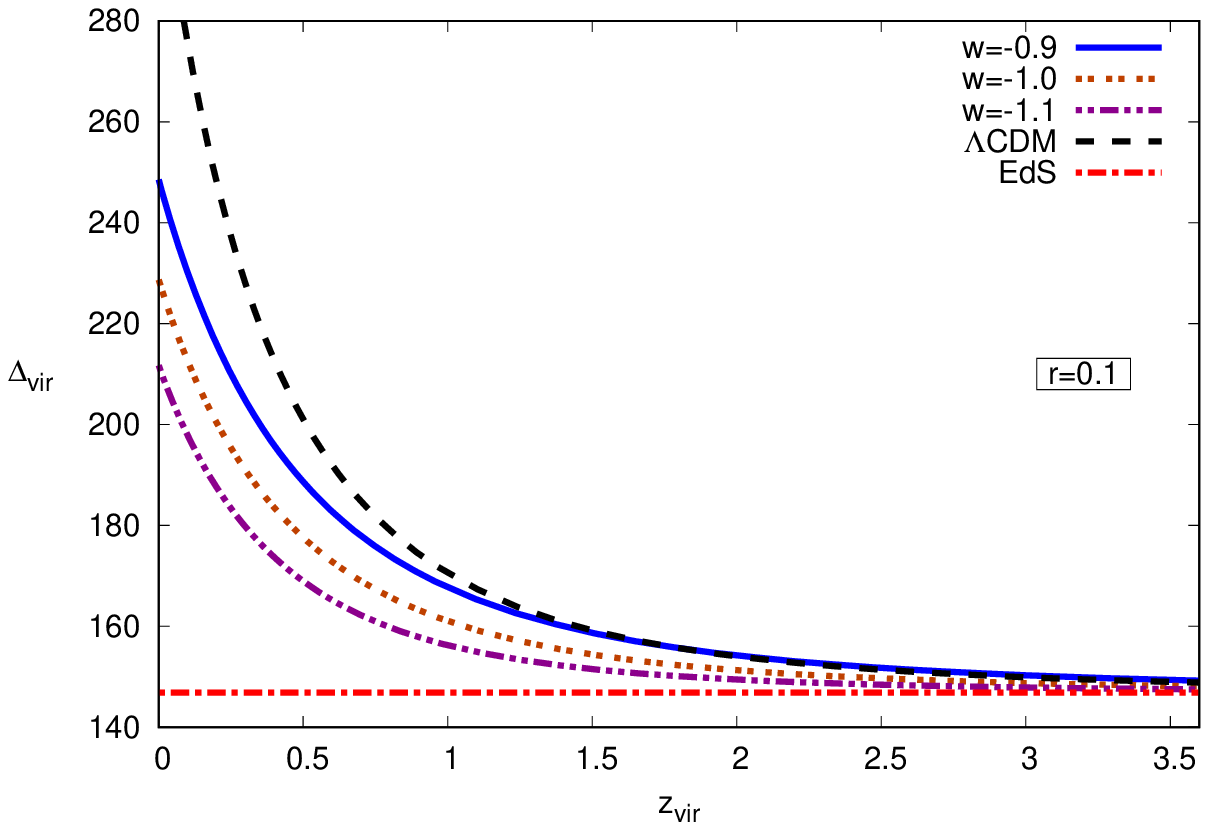}
\label{fig:figure2}
\end{minipage}
\caption{The nonlinear overdensity $\D_{\vir}$ for a couple of DE models with $r=1$ (upper panel) and $r=0.1$ (lower panel). In each plot, the $\Lambda$CDM and the EdS universe are also shown. For cases with a greater $w$, the structure is formed earlier and $\D_{\vir}$ gets larger at the same redshift. All of them converges to the minimum $\D^{\rm EdS}_{\vir}\approx146.8$ for early virialization.}
\end{figure}

One can also determine the nonlinear overdensity $ \D_{\vir} $ at the epoch of virialization by means of
\begin{equation}
\label{D}
\D_{\vir}\equiv\left.\f{\mc}{\mb}\right|_{z_{\vir}}
=\zeta\left(\f{x_{\vir}}{y_{\vir}}\right)^3,
\end{equation}
which is portrayed as a function of $z_{\vir}$ in Fig. 4 for a couple of DE models with $r=1$ and $0.1$. Our calculation shows that $ \D_{\vir}^{\rm{EdS}}\simeq146.8 $ for the EdS universe, which is consistent with previous results~\cite{sky,yywong}. The DE model with a greater $w$ leads to a larger $ z_{\vir} $, which implies an earlier structure formation. Consequently, it ends up with a greater nonlinear overdensity. Without the contribution from the DE-induced potential $\Phi_{\de}$, however, the $\Lambda$CDM generally reaches the highest $\D_{\vir}$. In stark contrast with the $\Lambda$CDM, the EdS universe only attains the minimum nonlinear overdensity due to a lack of repelling gravitation.

The virialization is certainly an important step toward the realization of the actual process of structure formations. There exists many viable but different approaches. No standard scenario has yet been formulated. According to Eqs.~\eqref{virial} and~\eqref{U} we know that the potential energy of the system, thus the DE potential, and the initial condition at the onset of recollapse play crucial roles in the process. Here we have used the parameter $r$ to characterize the effective energy density of DE within the overdense region at the turnaround epoch while keeping the eos and $\Lambda$ constants. Under the circumstances, it has been shown that, in addition to the specific form of the DE potential being critical to the outcome of the virialization, the clustering and non-clustering cases may not be too different as long as the eos does not deviate too much away from $w=-1$~\cite{motabruck}.  On the other hand, homogeneous and inhomogeneous vacuum energy with time-varying eos have also been considered, as in~\cite{voglis, bps}. It is conceivable that the end product of the virialization depends strongly on the choice of the vacuum energy. Regardless the higher amplitude in $\D_{\vir}$, their results show that values of $y_{\vir}$ in all different kinds of DE models approach $0.5$ while tending to the EdS universe at high redshifts, which are consistent to what we have presented in this section.


As mentioned previously, the clustering DE for the spherical overdensity ($\alpha$=0 in Eq. \eqref{dec}) results in an isolated system sitting in the otherwise homogeneous background universe. The energy within the overdense region does not leak out in the whole process of collapsing.  Following Ref.~\cite{lahav}, we have analyzed its virializing state by letting the whole system virialize (while only matter virializing for $\Lambda$CDM) and utilized the virial theorem and energy conservation for the virialization process in this extremely inhomogeneous DE case. Our results show that the values of $y_{\vir}$ for the whole system virialization in all DE models are always larger than $0.5$ while for $\Lambda$CDM it is lower than $0.5$, which are consistent with the results in Ref.~\cite{lahav}. However, for the case that DE is not extremely clustering ($\alpha\neq 0$), the DE component within the spherical overdense region does not conserve energy. In fact, the system is no more isolated and the energy leaks to the background universe. Under this circumstance, one needs to take the energy loss into consideration. The authors in Ref.~\cite{lahav} have corrected for this energy conservation problem by introducing the adjustable parameter $\alpha$ with a value between $0$ and $1$ representing different clustering behavior of DE. In the case of homogeneous DE, for example with $\alpha=1$, one can assume that the DE density is time independent; that is, the DE density at virialization is the same as that at turnaround~\cite{sky,pwang06}, where it was showed that the values of $y_{\vir}$ and $\D_{\vir}$ approach to the ones of EdS universe at high redshifts, being similar to our results. It is apparent that the results for $y_{\vir}$ and $\D_{\vir}$ are not so different in the clustering and non-clustering cases. It should be noted that when deriving the $y_{\vir}$ Eq.~(\ref{eq_yvir_m}), we have used the conservation of the total mass of matter inside the spherical overdensity. However, one can further explore the effect of clustering DE on the halo mass~\cite{crem,yywong2} where the total mass may not be conserved. In particular, the authors in Ref.~\cite{yywong2} have studied the virialization condition by considering mass non-conservation with arbitrary DE sound speed, between the clustering and non-clustering limits. As a consequence, it leads to distinct results from what we have obtained. For example, their values of $y_{\vir}$ for different DE models are all lower than $0.5$.

\section{complete journey of clustered matter density perturbation}

We have worked out the nonlinear evolution of overdensity within a spherical region filled with clustered DE during the process of recollapsing. However, it is intriguing to explore the prehistory of those fully clustered density perturbations prior to the turnaround epoch. Combining the linear perturbation approach, the SCM actually allows us to trace out the complete journey of the matter density contrast right from very early times to the settlement of large scale structures.

Since the gauge issue is negligible at scales much less than the Hubble radius, the Newtonian formulation~\cite{lima} is sufficient to describe the motion of density contrasts.  Expressing in terms of comoving coordinates, the evolution of density contrasts in matter, $\dm=\mc/\mb-1$, and DE, $\dde=\dec/\deb-1$, in the linear regime is governed by the perturbed continuity equations,
 \begin{eqnarray}
 \label{pcm}
 \dot\delta_\m+{1\over a}\overrightarrow{\nabla} \cdot \vec{u}&=&0, \\
 %
 \label{pcx}
 \dot\delta_\de+{1\over a}(1+w)\overrightarrow{\nabla} \cdot \vec{u}&=&0,
 \end{eqnarray}
where $\vec{u}$ denotes the peculiar velocity of matter and we have assumed that $\delta P_{\de}=w\delta\dec$; the perturbed Euler equation,
%
\begin{equation}
\label{euler}
\dot{\vec{u}}+H\vec{u}=-{1\over a}\overrightarrow{\nabla}\delta\Phi,
\end{equation}
where $\delta\Phi$ is the perturbed gravitational potential; and the Poisson equation,
%
\begin{equation}
\label{poisson}
\overrightarrow{\nabla}^2\delta\Phi=4\pi G\left(\delta\rho_\m+\delta\rho_\de+3\,\delta\hspace{-.2mm} P_\de\right)a^2.
\end{equation}
With the help from Eqs.~\eqref{euler} and \eqref{poisson}, the continuity Eqs.~\eqref{pcm} and \eqref{pcx} illustrating the dynamics of the cosmic fluid can be rewriten as  
%
\begin{equation}
\label{dm_0}
\ddot{\delta}_{\m}+2H\dot{\delta}_{\m}
=4\pi G(\mb\dm+\deb\dde+3w\deb\dde),
\end{equation}
and
\begin{equation}
\label{dde_0}
\dot{\delta}_{\de}=(1+w)\dot{\delta}_\m.
\end{equation}

Following the same strategy as in Sec.~II, we solve Eqs. \eqref{dm_0} and \eqref{dde_0} numerically by adopting the normalized scale factor $ \x $ and the cosmic time $ \eta $ such that
%
\begin{eqnarray}
\x   &\equiv& \f{a}{a_0}, \\
\eta &\equiv& \sqrt{\Om_{\m}^0}H_0t,
\end{eqnarray}
then the Friedmann Eq. \eqref{H} becomes
\begin{equation}
\label{dx'}
\f{d\x}{d\eta}=\f{1}{\sqrt{\x\Om_{\m}(\x)}},
\end{equation}
in which 
\begin{equation}
\Om_{\m}(\x)=\left(1+\f{1-\Om_{\m}^0}{\Om_{\m}^0}\x^{-3w}\right)^{-1}.
\end{equation}
Accordingly, the dynamical Eqs. \eqref{dm_0} and \eqref{dde_0} can be recast as
%
\begin{equation}
\label{dm}
\f{d^2\dm}{d\eta^2}+\f{2}{\x}\f{d\x}{d\eta}\f{d\dm}{d\eta}
=\f{3}{2\x^3}\left[\dm+\f{1-\Om_{\m}^0}{\Om_{\m}^0}(1+3w)\x^{-3w}\dde\right],
\end{equation}
%
\begin{equation}
\label{dde}
\f{d\dde}{d\eta}=(1+w)\f{d\dm}{d\eta}.
\end{equation}
Since all non-adiabatic perturbations are strongly constrained by CMB data, it is reasonable to presuppose that non-adiabatic modes of density perturbation must decay away as the universe expands~\cite{abramo07}. Under the circumstances, Eq. \eqref{dde} can be simplified as
\begin{equation}
\label{dde-dm}
\dde=(1+w)\dm.
\end{equation}

We assume that the matter density contrast $\dmi$ emerges at some early moment of $\eta_i$ in the matter dominant universe, so that $\ddei=(1+w)\dmi$ and $\x_i=(3\eta_i/2)^{2/3}$. The presumption of linear growing at the early stage, i.e. $\dm\propto a$, leads to $\dot{\delta}_{\m}=H\dm$, which in turn provides the initial growth rate of $\dm$ at $\eta_i$, i.e. $d\dm/d\eta|_i=2\delta_{\rm{m},\it{i}}/(3\eta_i)$. The history of the matter density contrast is then unfolded according to Eqs.~\eqref{dx'} and~\eqref{dm}. Tracing out the full evolution toward the formation of structures, however, we need to determine the ending moment $\eta_c$ of the whole recollapsing process. This is where SCM comes into play~\cite{crem}.

To begin with, we rescale the radius $R$ of the overdense region to unity at the initial time such that the size of the overdensity is measured by the dimensionless quantity $\y$, i.e.
\begin{equation}
\y\equiv\f{R}{R_i}.
\end{equation}
Subsequently, Eq. (5) monitoring the development of the overdense region filled with a clustered DE ($\alpha=0$) can be rewritten as
\begin{equation}
\label{dy'}
\f{d^2\y}{d\eta^2}+\f{1}{2}\left[
\f{1+\dmi}{\x_i^3}\f{1}{\y^2}+(1+3w)(1+\dde^{\rm{NL}})
\f{1-\Om_{\m}^0}{\Om_{\m}^0}\f{\y}{\x^{3(1+w)}}\right]=0,
\end{equation}
where Eqs. \eqref{mc} and \eqref{dec} have been used. The variation in the nonlinear density contrast of DE, i.e. $ \dde^{\rm{NL}}$ as defined in Sec.~II, is characterized by Eq. \eqref{dec} which can be further modified as
\begin{equation}
\label{dde_NL}
\f{d\dde^{\rm{NL}}}{d\eta}+3(1+w)(1+\dde^{\rm{NL}})
\left(\f{1}{\y}\f{d\y}{d\eta}-\f{1}{\x}\f{d\x}{d\eta}\right)=0,
\end{equation}
according to Eq. \eqref{deb}. Similarly with the help of Eq. \eqref{mb}, the change in the nonlinear matter density contrast $\dm^{\rm NL}$ guided by Eq. \eqref{mc} can be reworded as
\begin{equation}
\label{dm_NL}
\f{d\dm^{\rm{NL}}}{d\eta}+3(1+\dm^{\rm{NL}})
\left(\f{1}{\y}\f{d\y}{d\eta}-\f{1}{\x}\f{d\x}{d\eta}\right)=0.
\end{equation}

In this approach, we treat both $\dm^{\rm NL}$ and $\dde^{\rm{NL}}$ as linear quantities at early times in the matter-dominated era. Thus, $\ddei^{\rm{NL}}\approx\ddei=(1+w)\dmi$ for a given $\dmi$. On the other hand, since $\dm\propto\x\propto\eta^{2/3}$ in the linear regime, Eq. \eqref{dm_NL} can be further expressed as
\begin{equation}
\f{1}{\y}\f{d\y}{d\eta}=\f{2}{3\eta}\left(1-\f{1}{3}\f{\dm}{1+\dm}\right),
\end{equation}
which renders $d\y/d\eta|_i\approx2(1-\dmi/3)/(3\eta_i)$ as the initial velocity of recollapsing. We set $\eta_i=10^{-6}$ and $\Omega_\m^0 =0.274$~\cite{wmap5} for all our numerical calculations. 

\begin{figure}
\includegraphics[width=0.7\textwidth]{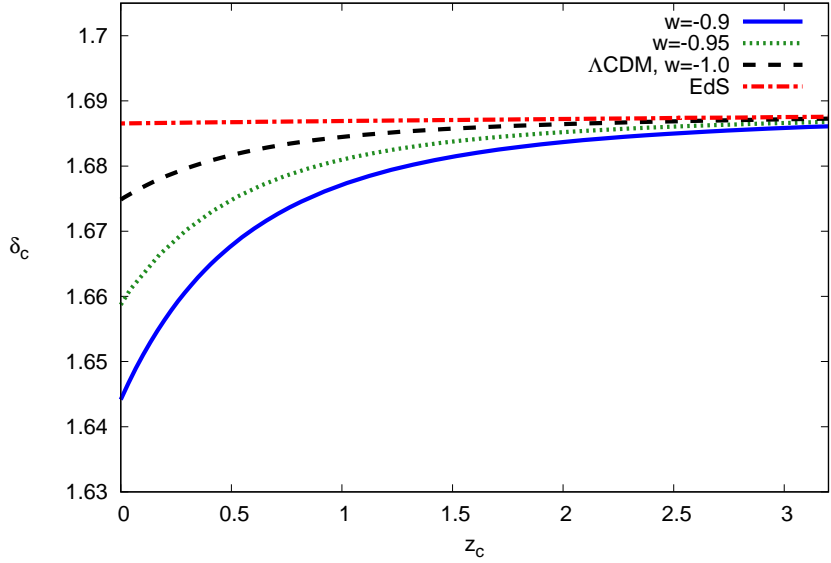}
\caption{The critical density contrast at collapse $\dc$ versus the redshift at collapse $z_{\rm c}$.}
\end{figure}

Being the matter density contrast at collapse calculated in the linear regime, i.e. $\dc=\dm(\etac)$, the critical density contrast $\dc$ can now be determined through the following procedure. Given a set of proper values for initial conditions and $w$, we solve Eqs.~\eqref{dx'},~\eqref{dy'} and~\eqref{dde_NL} at the same time to get $\etac$ which is identified as the moment pining down the epoch of $\y=0$. Adopting $\etac$ as the upper limit, $\dc$ is obtained by integrating Eq.~\eqref{dm} out with the help of Eq.~\eqref{dde-dm}. We plot the results for a couple of clustered DE models, $ \rm{\Lambda CDM} $, and the EdS universe in Fig. 5. Apparently, $ \dc $ at high redshifts converge to the well-known $ 1.686 $~\cite{sky}, which is the standard value in the EdS universe.

Following a similar scheme, one can also compute the critical density contrast at the virialized time, i.e. $\delta_\vir=\dm(\eta_\vir)$. After solving Eqs.~\eqref{dx'},~\eqref{dy'},~\eqref{dde_NL} and ~\eqref{dm_NL} simultaneously, the turnaround time is located at the moment $\eta_\ta=\etac/2$. Subsequently, the density ratio $q$ of DE to matter at turnaround can be obtained by its original definition. Applying the constraint Eq.~\eqref{eq_yvir}, it is straightforward to get $\eta_\vir$ which is then served as the upper limit to integrating out Eq.~\eqref{dm} and determining $\delta_\vir$. For example, Fig. 6 dipicts several outcomes of our calculations. The value of $ \dvir $ for the case of EdS universe agrees with that acquired in Ref.~\cite{sky}. Though the difference between the DE model with $w=-1$ and $\Lambda$CDM is clearly discernible, the regular pattern of evolving trend for $\dc$ with respect to $w$ as shown in Fig. 5 is severely spoiled by the process of virialization as shown with $\dvir$ in Fig. 6.
%
\begin{figure}
\includegraphics[width=0.7\textwidth]{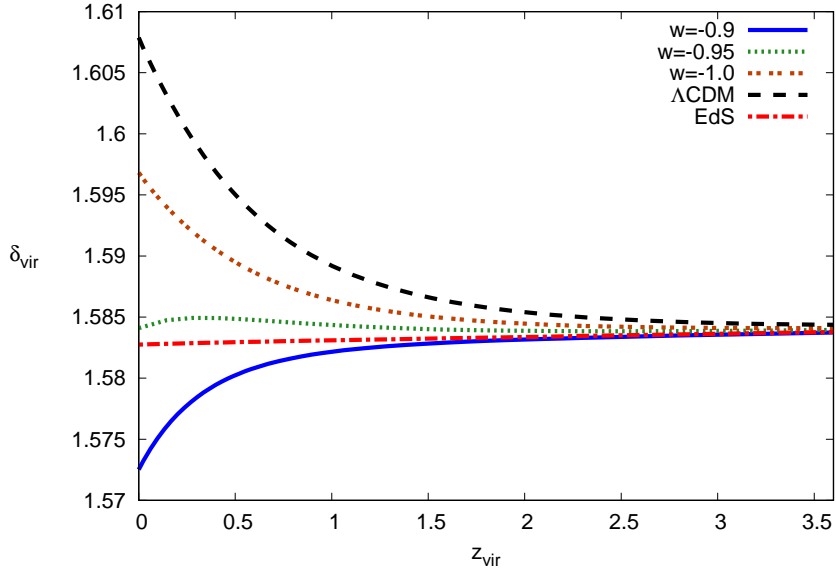}
\caption{The linear perturbation at the virialization time versus the redshift of virialization.}
\end{figure}

\begin{figure}
\includegraphics[width=0.7\textwidth]{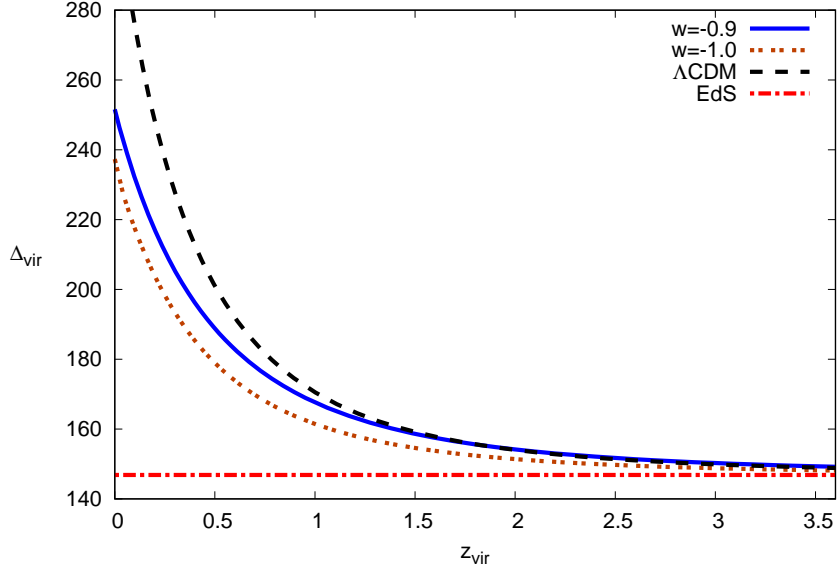}
\caption{Nonlinear overdensity $ \D_{\vir} $ at virialization determined by Eq.~\eqref{Dvirlin} for a couple of clustered DE models and $ \rm{\Lambda CDM} $. Their behavior is similar to those shown in Fig. 4. The range of initial matter density contrast that we have used to produce the curves is  $2.9\times 10^{-4} \leq \dmi \leq 9.6\times 10^{-4}$.}
\end{figure}
Moreover, the entire course of evolution for nonlinear overdensity is trackable if we assume that large-scale structures are indeed originated in the initial matter density contrast $\dmi$ at early times. The virialized nonlinear overdensity defined in Eq. \eqref{D} can be recast as
\begin{equation}
\label{Dvirlin}
\Delta_{\rm{vir}}=\left(1+\delta_{\rm{m,ta}}^{\rm{NL}}\right)\left(\frac{a_{\rm{vir}}/a_{\rm{ta}}}{R_{\rm{vir}}/R_{\rm{ta}}}\right)^3.
\end{equation}
Accordingly, given a $\dmi$, the corresponding $\Delta_\vir$ can be obtained under the constraint of Eq.~\eqref{eq_yvir} by solving Eqs.~\eqref{dx'} and \eqref{dy'}-\eqref{dm_NL}. Therefore, it is straightforward but tedious to trace out the complete evolution of the clustered mass perturbation in a particular clustered DE model with a fixed eos $w$ by repeating the above mentioned procedure. The evolving trend of the solutions for the nonlinear overdensity $\Delta_\vir$ determined by Eq.~\eqref{Dvirlin} with different $w$ is shown in Fig. 7, which is similar to those shown in Fig. 4.     
The ending moment $\etac$ of spherical collapse obviously plays the critical role in finding matter overdensities $\dc,\ \dvir$, and $\Dvir$. It is closely linked with the initial matter density contrast $\dmi$. As a consequence, the recollapsing of density perturbations, thus the formation of large-scale structures, is firmly built upon the amplitude of $\dmi$. Our numerical experiments have shown that the linear matter perturbation $\dm$ would generally escape from recollapsing whenever $\dmi < 2.2\times 10^{-4}$. To allow for recollapses in the DE models considered in Fig. 7, we have used initial matter density contrast within the range, 
$2.9\times 10^{-4} \leq \dmi \leq 9.6\times 10^{-4}$.

Indeed, the linear perturbation theory supplements the SCM with the growth of DE and matter perturbation in the linear regime before the non-linear process of recollapsing. But knowing the linear growth does not provide us with any additional information about the structure formation. The detailed process of recollapsing, however, depends on the values of the initial DE and matter perturbation. Furthermore, the whole recollapsing process spends relatively short time in the linear regime, during which or in an even earlier epoch the DE equation of state may not be the same constant value. As far as the formation of large-scale structures is concerned, we can trade the linear perturbation theory with a phenomenological parameter $r$, being the density ratio of matter to DE at turnaround time, as introduced in Eq.~\eqref{r} in the previous section. Given a value of $r$, the SCM is well defined and the SCM equations of motion carry legitimate solutions. Each of these solutions corresponds one-to-one to that in the linear perturbation theory. We have studied the correlation between the parameter $r$ and the initial amplitude of the matter density contrast $\dmi$ for different DE models in Fig. 8.  Accordingly, it is justifiable to properly recount the evolution of nonlinear matter density in terms of the parameter $r$. For example, the solution to $w=-0.9$ in Fig.~7 obtained by solving the full equations for matter and dark energy, including the case that $y_\vir = 0.5$, can be mimicked by the introduction of $r\simeq 0.04$ in the full non-linear equations of the spherical collapse model. We note that the correlation is less dependent of initial amplitude of the matter density contrast once $\dmi\ge 5\times 10^{-4}$, as illustrated in Fig. 8. 
%
\begin{figure}
\includegraphics[width=0.7\textwidth]{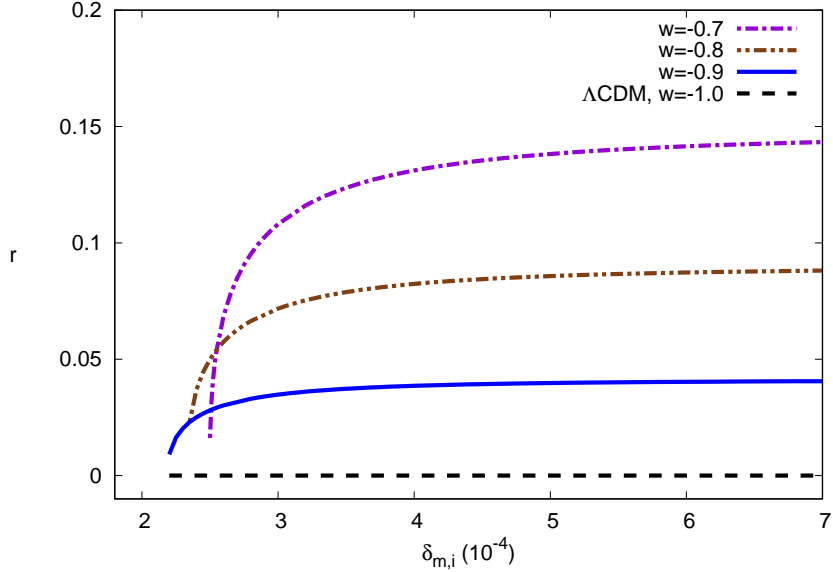}
\caption{The density contrast ratio of DE to matter at turnaround is plotted against the amplitude of initial density contrast. Apparently, the value of $r$ is less dependent of $\dmi$ as $\dmi\ge 5\times 10^{-4}$. }
\end{figure}

In fact, it is plausible to constrain the ratio $r$ using observational data on galaxy clusters. Considering a halo as a spherical overdense region of mass $M_\vir$, the virialized overdensity can be estimated as
%
\begin{equation}
\D_{\vir}=
\f{3M_{\vir}}{4\pi\rho_{\rm{c,0}}\Omega_{\rm{m},0}(1+z)^3R_{\vir}^3},
\end{equation}
where $ \rho_{\rm{c,0}} $ is the present critical density of the Universe. Using observational data of 7 clusters with projected axial ratio $>0.8$ that each include at least 6 galaxy members at an averaged redshift $\langle z\rangle\simeq0.015$~\cite{bps} in the Two Micron All-Sky Survey (2MASS) Extended Source Catalog~\cite{2mass}, the virialized overdensity at the present time is derived as $\Dvir(0) = 348 \pm 146$ at $2\sigma$ level. We have predicted the values of $\Dvir(0)$ from Eq.~\eqref{D} in the SCM for different clustered DE models with various $r$ and $w$, which are denoted by scattered symbols in Fig. 9. As a consequence, it is likely that $0.5 < r \leq 0.8$ for models containing clustered DE with $w\leq -0.9$ at a confidence level of $68\%$.
%
\begin{figure}
\includegraphics[width=0.7\textwidth]{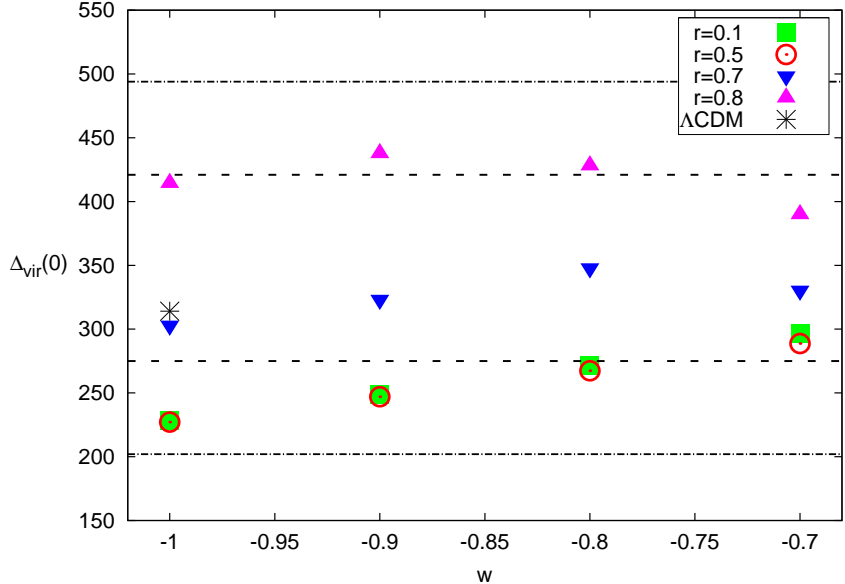}
\caption{Theoretical prediction of the virialized overdensity $\Dvir(0)$ in clustered DE models in terms of the ratio $r$ as a function of $w$, which are denoted by scattered symbols. The $ 1\sigma $ (the dashed line) and $ 2\sigma $ (the dot-dashed line) observational constraints on $\Dvir(0)$ are derived using a subgroup of galaxy clusters in the 2MASS High Density Contrast group catalog.}
\end{figure}

\section{Conclusions}

In virtue of the notion of SCM, we have investigated the clustering effect of DE on the evolution of matter density contrast in a spatially flat universe. With a fully clustered DE component, the spherical overdense region is considered as an isolated system behaving like a closed FLRW universe which conserves the energy separately for both matter and DE within it. As a consequence, the relative proportion in matter and DE components in the overdense region is underdetermined within the SCM. Other than using the theory of linear density perturbation that enables us to trace the early history of the matter and DE perturbation and thus relate them at a later time, we presuppose a linear relation that $ \nld_{\de,\ta}=r\nld_{\m,\ta} $ be hold for the matter and DE density contrasts at turnaround time to remove the difficulty with clustered DE in the SCM. We have shown that this simple relation can replace the complicated linear analysis, and allow the SCM to work efficiently and consistently especially in the epoch of structure formation. After pining down the time of complete collapsing, we have obtained the overdensity at turnaround $\tau_\ta$ as a monotonously decreasing function of turnaround redshifts $z_\ta$. That is, the later the spherical overdense region starts recollapsing, the more matter density it contains. Meanwhile, the growth in $z_\ta$ at the same turnaround redshift is proportional to the DE equation of state ($w$), due to the less repelling exercised by the DE with a greater $w$.  

Taking into account a proper scheme for virialization, the nonlinear overdensity $\Dvir$ as the end product of the recollapsing can be determined. Moreover, it is plausible to distinguish $\Lambda$CDM from the clustering model of DE with $w=-1$ according to whether it is the whole system or only the matter component that is virialized. Without the DE-induced potential in the process of virialization, $\Dvir$ in $\Lambda$CDM reaches the maximum value in contrast to all other clustered DE models, and the EdS universe. On the contrary, when the DE-induced potential becomes compelling during virialization, the amplitude of $\Dvir$ is proportional to $w$, which implies that the growth in overdensity is suppressed by the repelling of the DE component within the spherical region.   

With the help from the linear perturbation theory, it is straightforward but somehow more tedious to track down the complete evolution for the matter density contrast emerging at an early time toward the final moment of forming a stable large-scale structure within the boundary of the spherical overdense region. It turns out that the virialized overdensity $\Dvir$ obtained by this approach is in general closely related to the amplitude of initial matter density contrast $\dmi$. We have found that the criterion $\dmi \ge 2.2\times 10^{-4}\sim 2.5\times 10^{-4}$ depending on $w$ has to be fulfilled in order to form a cosmic structure. On the other hand, the ratio $r$ is less dependent on the initial matter density contrast for cases with $\dmi \ge 5\times 10^{-4}$. As a matter of fact, the numerical scheme with the ratio $r$ in solving for $\Dvir$ is a more efficient alternative to tracing out the entire course of evolution for the clustered matter density perturbation provided that $\dmi \ge 5\times 10^{-4}$. Moreover, the current observation on large-scale structure suggests that $0.5 < r \leq 0.8$ for models containing clustered DE with $w\leq -0.9$ at $1\sigma$ level.

%
\begin{acknowledgments}

We thank Seokcheon Lee for helpful discussions during the early development of this project. This work was supported in part by the Ministry of Science and Technology, Taiwan, ROC under Grants No. MOST104-2112-M-001-039-MY3 (K.W.N.) and No. MOST104-2112-M-003-013 (W.L.).

\end{acknowledgments}

%

%

\end{document}